\documentclass[12pt]{article}
\usepackage[utf8]{inputenc}
\usepackage[T1]{fontenc}
\usepackage{graphicx}
\usepackage{placeins} 
\usepackage{amsmath}
\usepackage{authblk}
\usepackage{caption}

\usepackage{geometry}
\usepackage{setspace}
\begin{document}
\title{Physics~of~unraveling~and~micromechanics~of hagfish~threads }
\author{
Mohammad Tanver Hossain$^{1,2}$, Dakota Piorkowski$^{3}$, Andrew~Lowe$^{3}$, Wonsik~Eom$^{1,4}$, Abhishek Shetty$^{5}$, Sameh~H.~Tawfick$^{1,2}$, Douglas S. Fudge$^{3}$, Randy H. Ewoldt$^{1,2}$}

\date{}  

\maketitle
\noindent$^{1}$Department of Mechanical Science and Engineering, The Grainger College of Engineering, University of Illinois Urbana-Champaign, Urbana, 61801, USA\\
$^{2}$Beckman Institute for Advanced Science and Technology, University of Illinois Urbana-Champaign, Urbana, 61801, USA\\
$^{3}$Schmid College of Science and Technology, Chapman University, Orange, 92866, USA\\
$^{4}$Department of Fiber Convergence Material Engineering, Dankook University, Yongin-si, Republic of Korea\\
$^{5}$Rheology Division, Anton Paar USA, 10215 Timber Ridge Dr, Ashland, VA, 23005, USA
\begin{abstract}
\footnotesize
Hagfish slime is a unique biological material composed of mucus and protein threads that rapidly deploy into a cohesive network when deployed in seawater. The forces involved in thread deployment and interactions among mucus and threads are key to understanding how hagfish slime rapidly assembles into a cohesive, functional network.  Despite extensive interest in its biophysical properties, the mechanical forces governing thread deployment and interaction remain poorly quantified. Here, we present the first direct in situ measurements of the micromechanical forces involved in hagfish slime formation, including mucus mechanical properties, skein peeling force, thread–mucus adhesion, and thread–thread cohesion. Using a custom glass-rod force sensing system, we show that thread deployment initiates when peeling forces exceed a threshold of approximately 6.8~nN. 
To understand the flow strength required for unraveling, we used a rheo-optic setup to impose controlled shear flow, enabling us to directly observe unraveling dynamics and determine the critical shear rate for unraveling of the skeins, which we then interpreted using an updated peeling-based force balance model.
Our results reveal that thread–mucus adhesion dominates over thread–thread adhesion and that deployed threads contribute minimally to bulk shear rheology at constant flow rate. These findings clarify the physics underlying the rapid, flow-triggered assembly of hagfish slime and inform future designs of synthetic deployable fiber–gel systems.

\end{abstract}




\normalsize

\section{Introduction}

\begin{figure*}[t]
    \centering
    \vspace{-0.5cm}
    \includegraphics[width=\linewidth]{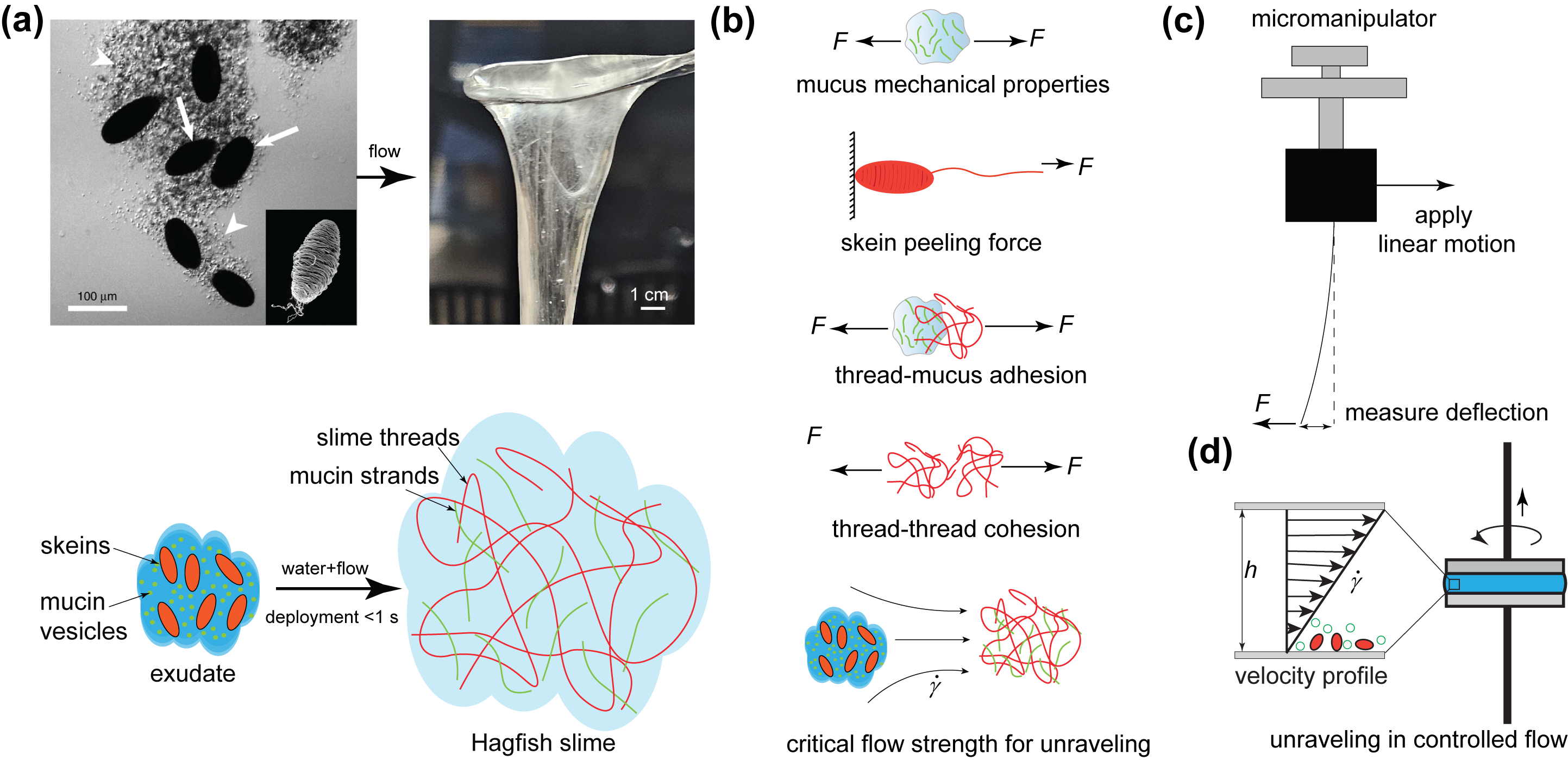}
    \caption{Underlying physics of hagfish thread unraveling and interactions.
(a) Optical image and schematic of hagfish exudate showing thread skeins (arrows), mucin vesicles (arrowheads), and inset of thread in its initial conical loop arrangement (height of the skein is 120 $\mu$m) (adapted from \cite{1,2,3}); alongside an image of hagfish slime formed by mixing exudate with seawater. A schematic below illustrates the components involved in slime formation, where hydrodynamic forces from the surrounding flow contribute to skein unraveling and network formation. The resulting threads, together with mucus, form a cohesive network that entrains a large volume of water.
(b) Schematic overview of the key mechanical interactions investigated in this study: (from top to bottom) mucus mechanical properties, skein peeling force, thread--mucus adhesion, and thread--thread cohesion. These factors collectively determine the critical flow strength required for slime deployment.
(c) Micromechanical measurement setup using a glass rod actuated by a micromanipulator to apply linear motion and quantify the deflection to estimate the forces involved.
(d)~Parallel-plate rotational rheometer setup used to impose controlled shear flows with optical access to evaluate skein unraveling behavior under defined flow conditions. }
    \vspace{-0.5cm}
    \label{fig:start}
\end{figure*}

Hagfish slime is a remarkable biological material that consists primarily of two distinct components (Figure \ref{fig:start}a): mucus produced by specialized gland mucous cells, and fibrous protein threads produced by gland thread cells (GTC) \cite{1,5,6,7,8,9,10,11,12}. The basic properties of hagfish slime include its exceptional volumetric expansion ratio, capable of expanding rapidly up to approximately 10,000 times its original volume upon contact with seawater, forming one of the most dilute hydrogels known \cite{1,8,11,13,14,15,16,17,18}. This extreme dilution, combined with its cohesive and elastic nature, makes hagfish slime particularly intriguing from both biological and engineering perspectives \cite{10,19,20,21}. The properties of particular interest include the strength and stiffness of individual protein threads (approximately 180~MPa strength and 6.4 MPa stiffness), along with the ultra-soft elasticity of the slime network (linear elastic shear modulus of about 0.02 Pa) \cite{10,15}. When attacked by predators, hagfish slime deployment occurs rapidly, typically within less than 400 milliseconds~\cite{9,16,21}. Hagfish GTCs employ a unique biological packing strategy wherein the threads are meticulously coiled into highly organized conical loop arrangements within the cytoplasm of the cell~\cite{8,13}. Upon ejection, these coiled threads swiftly unravel and interact with mucin vesicles and seawater, forming the cohesive, mucus-like network (Figure \ref{fig:start}a and Figure~S1) \cite{16,19,23}.

Whether fluid flow can generate sufficient force to overcome the resistance to thread unraveling and drive the rapid deployment of hagfish slime has been the focus of extensive research aimed at elucidating the underlying mechanics of unraveling \cite{3,12,16,23}. Historically, Newby (1946) proposed an osmotic or explosive release mechanism~\cite{24}; however, this hypothesis has since been experimentally disproven. More recent studies, including Winegard \textit{et al.} (2010), emphasized that vigorous mixing with mucus is critical for initiating thread deployment \cite{7}. Although spontaneous unraveling of skeins has been reported \cite{12,25}, these studies do not fully explain the exceptionally rapid sub-second timescale of deployment.

A recent study by Chaudhary \textit{et al.} (2019) introduced a quantitative theoretical framework based on a force balance between the fluid drag force pulling on the thread and the peeling force resisting thread release \cite{16}. This competition is captured by the dimensionless peeling number, $\wp = F_{\text{drag}}/F_{\text{peel}}$. When the fluid drag surpasses the resisting peeling force (\(\wp > 1\)), unraveling proceeds rapidly, approaching a kinematic limit where the surrounding flow directly advects the thread.

To date, experimental measurements of peeling force values have been lacking, introducing significant uncertainty into the unraveling model. Precise knowledge of these forces in situ would greatly enhance model validation. Furthermore, while previous work has focused on bulk deployment dynamics or spontaneous unraveling, little is known about the specific mechanical resistance that threads encounter during the transition from skeins to a fibrous network. These forms of resistance include the peeling force required to unravel threads from the skein (potentially arising from the biological glue holding the skeins) and adhesion forces among threads, both of which slow down the unraveling (Figure~\ref{fig:start}b). Moreover, adhesive interactions with the surrounding mucus and its rheology may modulate how closely thread motion follows the local flow. Strong thread-mucus adhesion may promote advection with the surrounding flow, while a stiff mucus network can compromise the softness of the resulting slime network.  Quantifying these forces can guide the understanding of the critical flow strength to initiate unraveling (Figure~\ref{fig:start}b).

To address this, we report for the first time direct experimental measurements of the in situ mechanical forces governing hagfish slime deployment, including the mechanical properties of mucus, the peeling force of hagfish threads, thread–mucus adhesion, thread–thread adhesion, and critical flow strength for unraveling. Using a custom-designed glass rod apparatus, we selectively attached to individual threads or mucus regions within the exudate and applied controlled separation speeds to quantify these forces (Figure \ref{fig:start}c). We hypothesize that an adhesive mechanism maintains skein integrity and that successful unraveling requires overcoming this adhesive resistance. Our measurements show an initial force required to detach a thread from the skein, which decreases as additional thread is pulled, producing a characteristic sawtooth force–extension profile. This suggests that thread deployment involves overcoming localized mechanical barriers, possibly due to adhesive contacts or coiling architecture within the skein. We find that thread–thread adhesion forces are comparatively negligible and unlikely to impede unraveling, while thread–mucus adhesion forces are substantially higher than the peeling force, indicating that mucus may play a dominant role in transmitting unraveling forces. Additionally, the mucus network exhibits a high extensional strain-to-break, enabling large deformations that likely help maintain the extreme softness and extensibility of the fully formed slime. 

To understand the flow strength required for unraveling, we used a rheo-optic setup to impose controlled shear flow on bulk skeins with mucin vesicles~(Figure~\ref{fig:start}d) surrounded by aqueous buffer solution with viscosity, $\eta\approx3$~mPa.s. This approach enabled us to directly observe unraveling dynamics and determine the critical shear rate for unraveling of the skeins, which we then interpreted using an updated peeling-based force balance model.
We observed that skein unraveling consistently occurred in simple shear flow at shear rates between 25 and 100~s$^{-1}$ (shear stresses 0.08 Pa to 0.3 Pa). By incorporating our in situ peeling force measurements into a previously proposed unraveling model \cite{16}, we predicted a critical strain rate of approximately 18~s$^{-1}$, in reasonable agreement with experimental shear flow results. Bulk skeins in the absence of a mucus network also unraveled at approximately the same critical flow strength, indicating that the presence of mucus is not necessary to initiate deployment. Furthermore, trapping a skein in the stagnation plane of counter-rotating plates did not reduce the unraveling threshold, suggesting that free skeins do not unravel more easily than pinned ones. These results confirm that hydrodynamic forces alone are sufficient to trigger thread deployment.




\begin{figure*}[t]
    \centering
    \vspace{-1cm}
    \includegraphics[width=\linewidth]{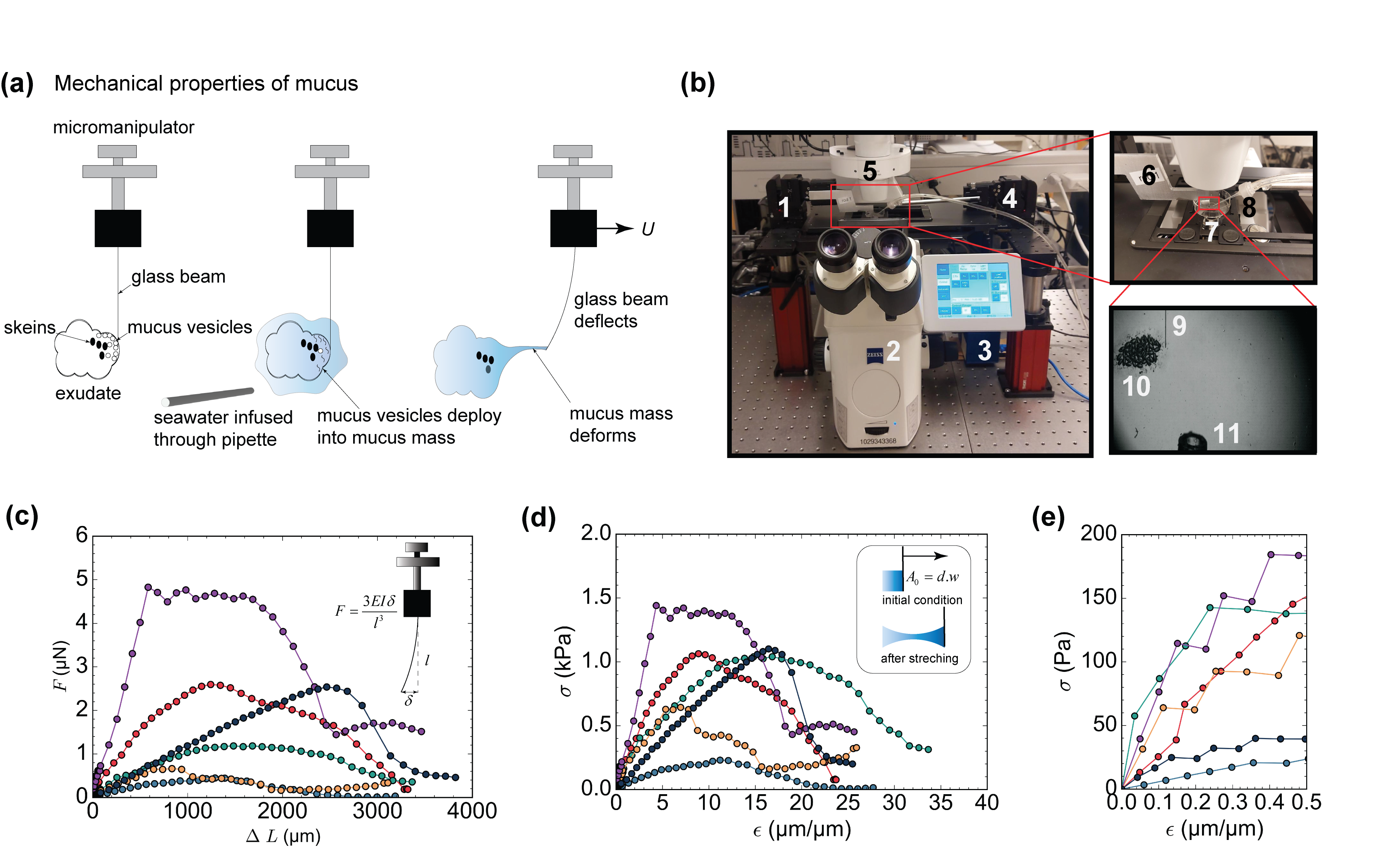}
    \caption{Mechanical characterization of hagfish slime mucus. 
    (a) Schematic of the experimental procedure to generate mucus from hydrated exudate and apply a lateral deformation using a calibrated glass beam to assess the mechanical response. 
    (b) Details of the experimental setup: An inverted microscope (2) with a high-speed video camera (3) enables imaging. A micromanipulator (1) holds the glass force-sensing rod, and a second micromanipulator (4) delivers artificial seawater via micropipette. The Petri dish (5) containing the exudate is positioned beneath the rod mounted on an aluminum holder (6), sample dish (7), and the water infusion nozzle (8). During testing, the glass rod (9) engages the mucus edge (10), hydrated via nozzle (11), and is displaced at a constant speed to induce deformation. 
    (c) Force-displacement profiles from individual trials showing peak load and extensibility. 
    (d) Converted engineering stress–strain curves reveal high extensibility; the inset shows schematics of the initial and stretched geometry of the mucus mass (see Table S1 for details).
    (e) Zoomed-in view of the small-strain region from panel (d) to measure the linear modulus across trials.}
    \vspace{-0.6cm}
    \label{fig:2}
\end{figure*}

\section{Methods}
\subsection{Animal care and slime gland exudate collection}

Pacific hagfish (Eptatretus stoutii) were collected from Monterey Bay (Chapman University). Animals were anesthetized with clove oil (150–200 mg/L) in 34\% artificial seawater (ASW), and slime exudate was collected using mild electrical stimulation (60 Hz, 1 ms, 18 V) as described in \cite{3,26}. The exudate, containing undeployed mucous vesicles and thread skeins, was collected with a Teflon-coated spatula and stored at 5$^{\text{o}}$C in stabilization buffer (0.55~M sodium citrate) \cite{27}. This buffer maintains skeins and vesicles in a condensed state, with seawater dilution triggering unraveling. Hagfish were returned to ASW after recovery. Mucus-only and thread-only samples were prepared by filtering the stabilized exudate through 40~µm cell strainers, which retain skeins and exclude mucous vesicles and were typically used for testing within two days of collection. 

\subsection{Micromanipulator setup}

To characterize the mechanical properties of hagfish slime mucus, skein peeling force, thread–mucus adhesion, and thread–thread adhesion under near-native conditions, we employed a custom-built glass beam mechanical testing system integrated with an inverted microscope and micromanipulator. Typically, the setup included a high-speed video camera for motion tracking, a syringe pump to infuse artificial seawater through a micropipette nozzle, and a micromanipulator holding a calibrated glass rod affixed to an aluminum rod using epoxy. Controlled displacement of the rod at 100 $\mu$m/s induced deformation in the subject, resulting in measurable bending of the glass rod unless otherwise mentioned. This bending was recorded with the high-speed camera and analyzed using the DLTdv8 MATLAB tracking tool to calculate the applied force from beam bending theory.  A more detailed description of the experimental setup, as well as the corresponding results, is provided in the results section.

\subsection{Flow setup}    

A MCR702 Twin Drive Rheometer was used (Anton Paar GmbH) to apply controlled shear flow conditions for unraveling experiments. A standard 43 mm diameter glass parallel plate was mounted to the upper and lower geometry to enable optical access during testing. The setup was integrated with a microscopy system, allowing simultaneous mechanical measurements and high-resolution visualization of skein deformation and unraveling dynamics under shear~\cite{28}.

\section{Results}

\begin{figure*}[t]
    \centering
    \includegraphics[width=\linewidth]{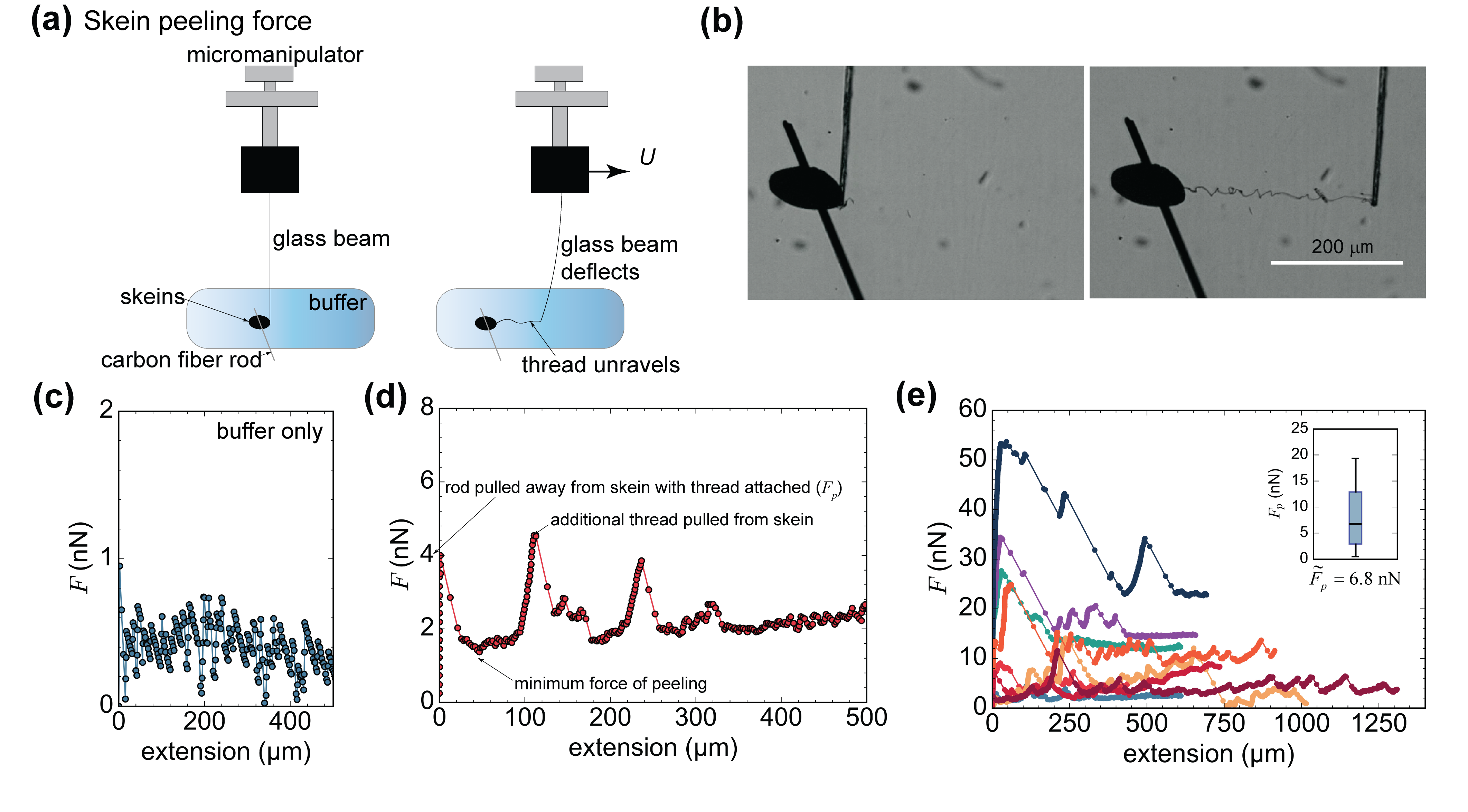}
    \caption{Characterization of hagfish skein peeling force.
(a) Schematic of the experimental procedure to measure the peeling force required to unravel hagfish skein threads. A calibrated glass beam attached to a micromanipulator pulled threads from skeins immobilized on a carbon fiber rod submerged in buffer solution.
(b) Representative optical images showing a thread attached between the carbon fiber rod and glass rod during peeling.
(c) Control experiment showing force–extension behavior when only the buffer solution is present.
(d) Representative force–extension curve during skein peeling, showing an initial detachment force followed by sawtooth force patterns associated with additional thread unraveling.
(e) Force–extension curves from multiple independent peeling tests, each plotted with a different color. The inset shows the distribution of measured peeling forces, with a median value of 6.8~nN across trials.}
    \label{fig:3}
\end{figure*}

\subsection{Mechanical properties of mucus mass}

\begin{table*}[!htbp]
\centering
\caption{Summary of mechanical properties of hagfish mucus mass}
\begin{tabular}{l|c c c c}
\hline
 & \textbf{$E$} & \textbf{$\sigma_p$} &\textbf{$\varepsilon_p$} & \textbf{$\varepsilon_b$}  \\
 & (kPa) & (kPa) & ($\mu$m/$\mu$m)& ($\mu$m/$\mu$m)\\
\hline
\textbf{Average} & 0.18 & 0.92 &10.4& 27.1 \\
\textbf{SE}      & 0.05 & 0.17  & 1.9&1.4  \\
\hline
\label{tab:table1}
\end{tabular}
\end{table*}

\begin{figure*}[t]
    \vspace{-0.5cm}
    \centering
    \includegraphics[width=0.8\linewidth]{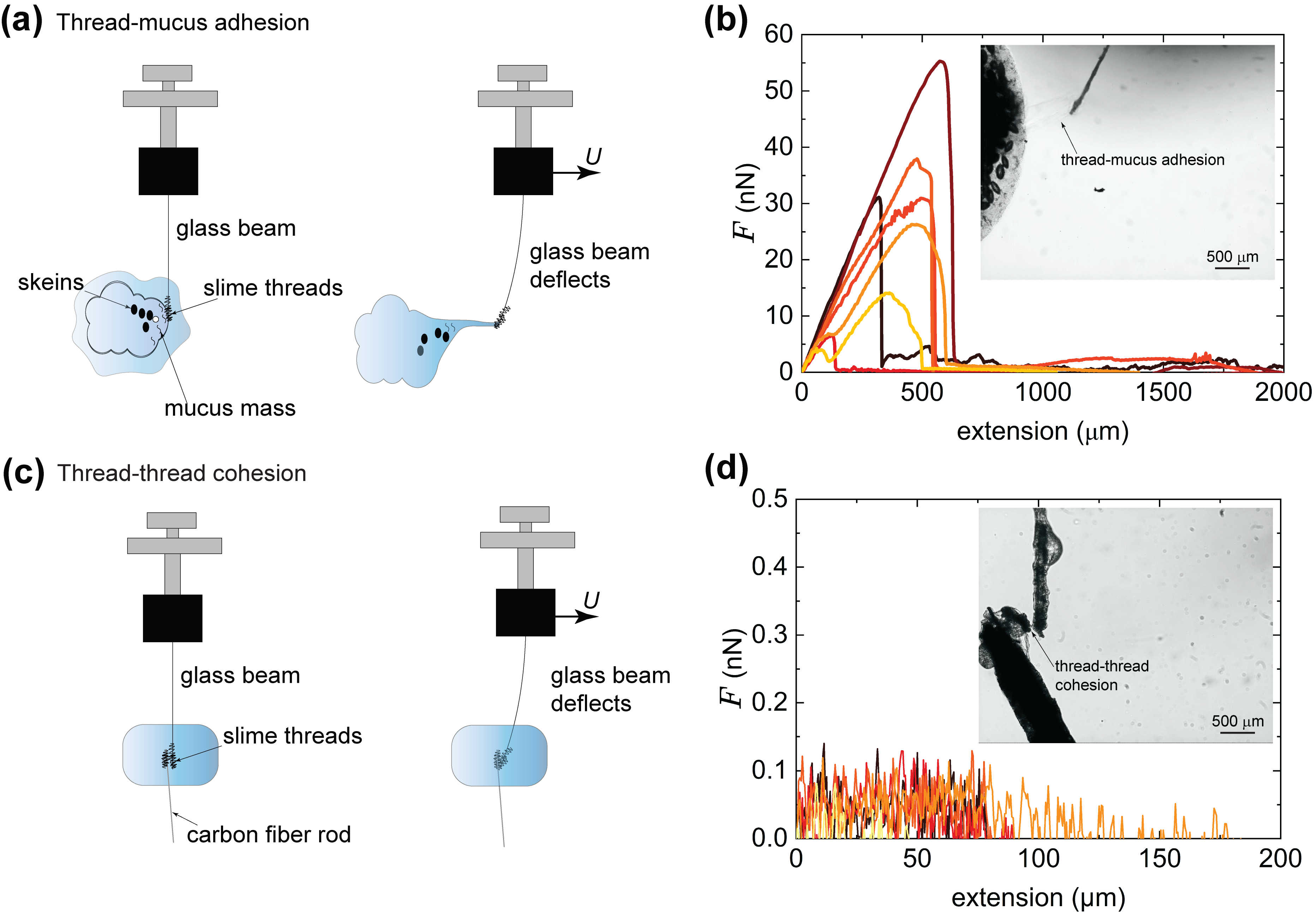}
    \caption{Characterization of thread–mucus adhesion and thread–thread cohesion forces.
(a) Schematic of the experimental procedure for thread–mucus adhesion. Threads unraveled from skeins were wrapped around a glass rod by twirling it in a drop of water containing dispersed threads. The rod was then brought into contact with a small blob of deployed gland exudate, and subsequently pulled away. The resulting deflection of the rod was used to infer the adhesive force.
(b) Representative force–extension curves showing the adhesive force between slime threads and the mucus mass. The inset shows an optical image of thread–mucus adhesion during testing.
(c) Schematic of the experimental setup for measuring thread–thread cohesion, with one set of threads immobilized by a carbon fiber rod and the others twirled around the end of a glass rod.
(d) Force–extension profiles for thread–thread cohesion tests, revealing small adhesive forces (<1~nN). The inset shows an optical image capturing thread–thread interaction during testing.}
    \label{fig:4}
    \vspace{-0.5cm}
\end{figure*}

To characterize the mechanical properties of hagfish slime mucus, the custom glass beam micromechanical testing system was employed, as illustrated in Figure \ref{fig:2}a. Freshly collected mucous vesicles and skeins from hagfish slime glands were deposited in a Petri dish and hydrated in situ by infusing artificial seawater through a micropipette. Upon hydration, mucus vesicles swelled and deployed into a continuous mucus mass. A calibrated glass beam, mounted on a micromanipulator, was brought into contact with the edge of the forming mucus. Controlled displacement of the rod at 100 µm/s induced deformation of the sample, resulting in measurable bending of the glass rod (Movie S1).  The viscous drag on the rod during motion was estimated to be 1~nN (see SI: viscous force estimation on slender rod). Thus, any measured force beyond this reflects true mechanical resistance from the mucus mass.

Details of the experimental setup used to measure the mechanical properties of hagfish slime mucus are shown in Figure \ref{fig:2}b. An inverted microscope equipped with a high-speed video camera provided real-time imaging of the deformation process. A micromanipulator was used to hold a calibrated glass rod, which served as the force-sensing element, while a second micromanipulator delivered artificial seawater through a micropipette. A close-up view shows the aluminum mount with the glass rod fixed using epoxy, positioned above the sample dish and aligned with the water infusion nozzle. During the experiment, the calibrated glass rod was brought into contact with the mucus edge of the hydrated exudate, which had been infused with artificial seawater. 

To quantify the force applied during mucus deformation and peeling experiments, we used multiple calibrated glass rods as a compliant beam with diameters $d$ and lengths $l$ of $(d,l)=$(10.85 $\mu$m, 1.7 mm), (13.97 $\mu$m, 2.8 mm), and (22.34~$\mu$m, 2.8 mm). The force \( F \) exerted on the rod tip is calculated from its deflection using the Euler-Bernoulli beam theory, assuming a cantilever beam fixed at one end. The relationship is given by \cite{29}
\begin{equation}
    F = \frac{3\delta EI}{l^3}
\end{equation}
where \( \delta \) is the deflection of the glass rod, \( E \) is the Young’s modulus of the glass, \( I \) is the second moment of area, and \( l \) is the length of the rod. The deflection \( \delta \) is determined as the difference between the displacement of the micromanipulator tip (\( x_m \)) and the actual displacement of the rod tip (\( x_r \)), i.e., \( \delta = x_m - x_r \). The Young’s modulus of the glass rod was independently measured as $E ~=~2.38~\times~10^{10}~\mathrm{N/m^2} $, determined from resonance-based calibration using the fundamental frequency of oscillation as \cite{30}
\begin{equation}
    f = \frac{1.875^2}{2\pi} \sqrt{\frac{EI}{\rho A l^4}}.
\end{equation}
Here, \( \rho \) is the material density, \( A= \pi d^2/4 \) is the cross-sectional area, and the second moment of area for the rod is given by \( I = \frac{\pi d^4}{64} \). This approach allows precise force measurements in situ.

Figure \ref{fig:2}c presents representative load-extension curves for hagfish mucus samples, obtained using the calibrated glass rod mechanical testing system. Each colored curve represents a different mucus sample collected from independent trials. The profiles exhibit non-linear force-extension behavior, typically featuring an initial stiff regime followed by a peak force and then a gradual decline. Maximum forces ranged between 0.5–5.5 $\mu$N, and the variability is likely due to varying amounts of mucus adhered to the glass rod in each trial. These curves were subsequently used to derive the corresponding engineering stress-strain curves (Figure \ref{fig:2}d). Engineering stress \( \sigma =F/A_0\) was computed by an initial cross section area comparable to the contact area \( A_0 = d \cdot w \), where \( d \) is the rod diameter and \( w \) is the contact width (obtained from video analysis of the mucus deflection trials and summarized in Table S1). Engineering strain \( \varepsilon \) was computed as \( \varepsilon = \Delta L / L_0 \), where \( \Delta L \) is the extension and \( L_0 \) is the initial mucus span (obtained from video analysis). The resulting curves reveal peak stresses \( \sigma_P \) on the order of kPa, and high extensibility, followed by pronounced strain-softening and, in many cases, a broad yielding plateau.

Four key parameters were extracted from the stress–strain response: Young’s modulus \( E \), peak tensile stress \( \sigma_p \), strain at peak stress \( \varepsilon_p \), and strain at break \( \varepsilon_b \). Young’s modulus was determined as the slope of the initial linear portion of the stress–strain curve. Strain at break, \( \varepsilon_b \), was calculated as the total strain at the point of rupture. Mucus samples typically showed soft modulus ($\sim0.18$~Pa), high extensibility (up to 2700\% strain), moderate peak strength ($0.92$~kPa), and a nonlinear stress response indicative of a yielding, soft elastic material. These properties are summarized in Table~\ref{tab:table1} with standard error (SE) across all samples.

\subsection{Skein peeling force}
Individual skeins were isolated to measure the force required to peel threads from hagfish slime skeins, and the resulting deflection was measured, as shown in Figure~\ref{fig:3}a. The forces involved in thread unraveling were significantly smaller than those required to deform the mucus mass. A longer and thinner glass rod ($l=5$5~mm, $d=8$~µm ) was used for more sensitive force measurement. To stabilize the skeins during testing, a shorter and stiffer carbon fiber rod ($l=2$~mm, $d=15$~µm) was positioned across the skein to pin it gently to the glass dish and prevent lateral movement.

Figure~\ref{fig:3}b provides images of the thread peeling experiment. Skeins suspended in 0.55~M sodium citrate buffer at 5$^\circ$C were injected into a glass dish on an inverted microscope and gently positioned using the glass and carbon fiber rods. After pinning the skein in place with the carbon fiber rod (oriented East–West relative to the camera frame), the tip of the long deflection glass rod was brought into contact with the distal end of the skein. To ensure attachment, the glass rod remained in contact with the skein for approximately 30 s. This passive contact method reliably established adhesion between the rod and the thread due to natural attraction, without requiring adhesives. Once attachment was achieved, the micromanipulator displaced the rod laterally by $\sim$1000~µm at a constant speed of 100~µm/s. The experiment was recorded at 50 frames per second, and rod deflection was subsequently tracked (Movie S2). 

Figure~\ref{fig:3}c shows the baseline force profile measured when the glass rod was moved through buffer solution in the absence of any skein or thread attachment. This control test was essential to quantify the background hydrodynamic resistance and system noise. The force–extension curve shows that the forces acting on the rod remained below 1~nN throughout the 500~µm displacement, with small fluctuations due to noise. Drag on a slender rod moving in a viscous fluid was estimated to cause deflection comparable to a 1~nN point load, which is consistent with our observations (see SI: viscous force estimation on slender rod, for details).  Thus, any measured force beyond this baseline during skein unraveling reflects actual mechanical resistance from thread peeling, rather than background drag forces.

Figure~\ref{fig:3}d presents a representative load–extension curve from a skein thread peeling experiment. The force profile illustrates two distinct stages: an initial peak corresponding to the detachment of the glass rod from the skein (defined as the onset of peeling), followed by a region of lower, fluctuating forces as the thread is continuously unraveled. The sawtooth-shaped force–extension profiles indicate an intermittent peeling mechanism rather than continuous detachment. This behavior is consistent with the release of discrete "hidden length" segments from the skein, likely influenced by variable coiling density. The force minima and maxima correspond to local thwarting and unthwarting events, rather than smooth elastic extension.

Figure~\ref{fig:3}e displays load–extension curves from multiple individual peeling tests, each represented by a distinct color. The initial peaks again reflect the higher force required to detach the rod from the bulk skein, while the subsequent lower-force regions capture single-thread unraveling dynamics. Force variation in this region arises from resistance due to thread entanglement, peeling out of the skein, and changes in curvature as the thread extends. The inset boxplot summarizes the distribution of peeling forces across all trials, yielding a median peeling force of approximately 6.8~nN and a standard error of 1.3~nN.

\begin{figure*}[!htbp]
    \centering
    \vspace{0cm}
    \includegraphics[width=\linewidth]{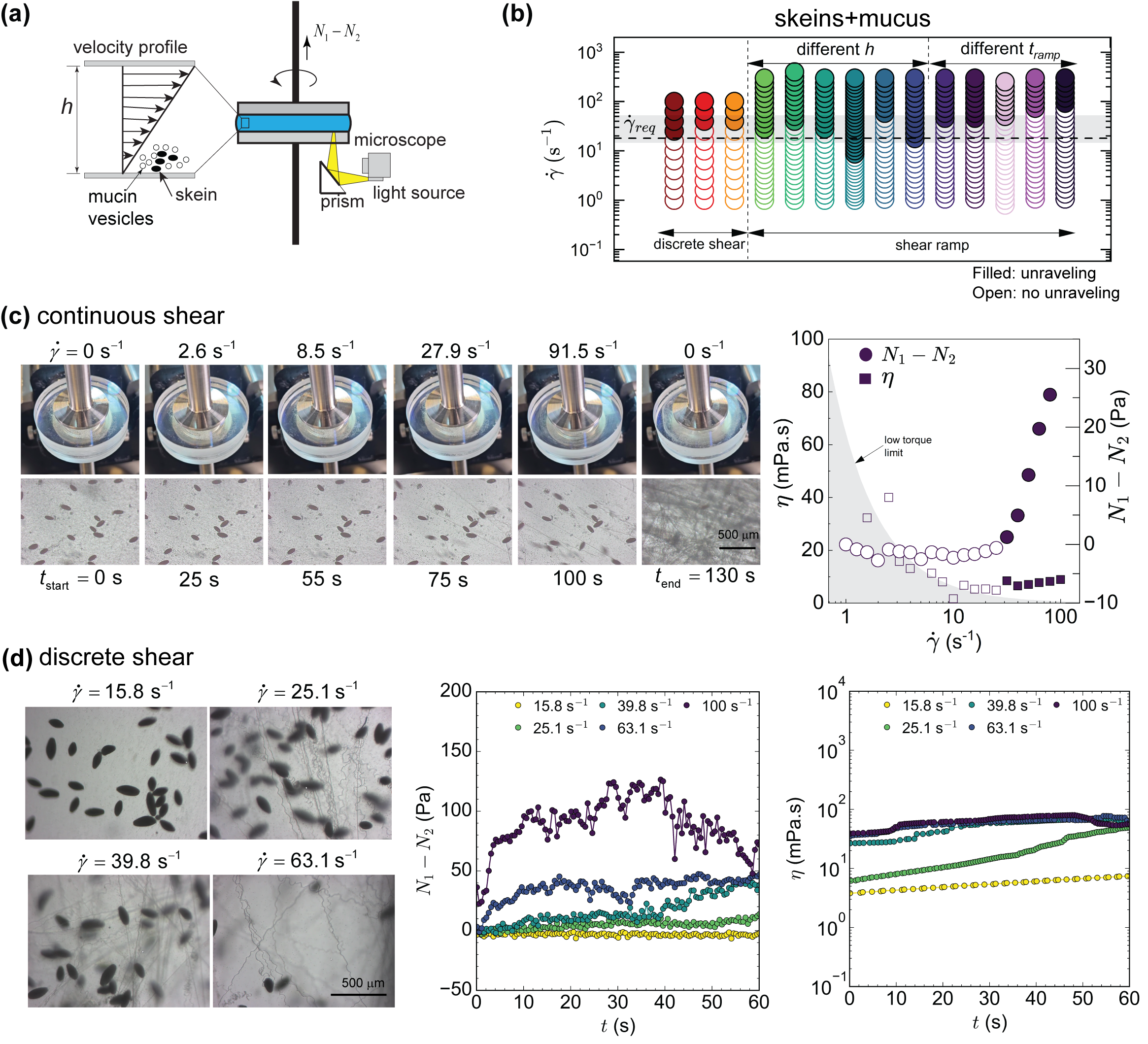}
    \caption{Characterization of skein unraveling dynamics and rheological behavior under controlled shear flow with mucin vesicles.
(a) Schematic of the experimental setup showing where skeins and mucin vesicles are added before shear.
(b) Unraveling map across different shear rates, with filled circles denoting observed unraveling, and open circles indicating no unraveling.
(c) Time-lapse images and rheological measurements during shear ramp experiments: at low shear rates ($\dot{\gamma} = 2.6$ to $8.5$~s$^{-1}$), minimal unraveling is observed, while at higher shear rates ($\dot{\gamma} = 27.9$ to $91.5$~s$^{-1}$), skein deployment was observed. The right panel shows steady-state viscosity and normal stress difference ($N_1-N_2$), with hollow symbols representing shear rates where no unraveling was observed and filled symbols representing shear rates where threads became optically visible.
(d)~Images (after 60 s of shear) and rheological response during discrete shear tests at different rates, revealing progressive increases in normal stress difference ($N_1-N_2$) and viscosity ($\eta$) associated with unraveling.}
    \label{fig:5}
    \vspace{-0.2cm}
\end{figure*}

\subsection{Thread-mucus adhesion force}    

Figure~\ref{fig:4}a illustrates the setup used to quantify the adhesive interaction between hagfish slime threads and the surrounding mucus. The mucus mass was formed in situ by adding a drop of freshly collected slime gland exudate and covering it with 95\% ethanol to keep it from drying out, but also keeping it from expanding too much while the glass rod force transducer was set up. In these trials, we used a glass rod force transducer that had been twirled within seawater containing unraveled slime threads, which resulted in the tip of the rod being wrapped in slime threads. We displaced the ethanol with seawater, which resulted in the swelling of the mucus, and approached the hydrated mucus with the thread-covered glass rod until contact was made, and then pulled it away at constant velocity using the micromanipulator. The resulting rod deflection captured the resistive force arising from thread–mucus adhesion (Movie~S3).

Figure~\ref{fig:4}b displays representative force–extension curves from multiple thread–mucus adhesion experiments, each shown in a different color. The inset micrograph shows the experimental geometry where the threads attached to the glass rod pull the mucus network, prior to adhesive failure. The force traces reveal a rapid buildup of tensile force as the adhered thread is displaced through the mucus, reaching peak values as high as 55~nN. The subsequent sharp drop in force indicates rupture of the adhesive contact between the thread and the mucus mass. In this case, $\epsilon_b$ is much smaller than that for the mucus component. 

Peak adhesion forces averaging 29~nN ($\pm$15nN), approximately 4–5 times greater than the typical 6.8~nN required for thread peeling from the skein, suggesting that mucus–thread adhesion is sufficiently strong to transmit unraveling forces during natural deployment.

\subsection{Thread-thread cohesion force}    

Figure~\ref{fig:4}c and \ref{fig:4}d illustrate the experimental approach and results for measuring cohesion forces between adjacent hagfish slime threads. A glass rod was prepared by twirling it within a small volume of seawater containing unraveled slime threads, which resulted in the end of the rod becoming wrapped in slime threads. Cohesion forces were measured by approaching a mass of immobilized threads with the glass rod until they made contact, which was observed as a slight bending of the glass rod. The glass rod was then pulled away using the micromanipulator. A portion of the slime threads was immobilized using a short carbon fiber rod. This setup mimics the scenario in which neighboring threads may interact during deployment. As the micromanipulator translated the rod at a constant velocity, any resistive force due to thread–thread adhesion would result in a measurable deflection of the calibrated glass beam (Movie S4).

Force–extension curves from multiple tests exhibited values consistently below 1~nN, i.e.\ at or near the resolution limit ($\sim0.4$~nN) of the glass rod system, as shown in Figure~\ref{fig:4}d. The inset figure confirms the interaction between neighboring threads, demonstrating that contact between threads resulted in very low adhesion forces. These results strongly suggest that thread–thread cohesion is negligible compared to thread–mucus adhesion or skein peeling forces, and thus likely plays a minimal role in the mechanical integrity or resistance during the formation of the slime network.

\subsection{Unraveling in controlled flow}

In this section, we directly observe unraveling dynamics and determine the critical shear rate required to initiate skein unraveling. 
Hagfish skeins were placed in a rheo-optical setup mounted on an Anton Paar MCR702e rheometer equipped with a separate motor-transducer arrangement. Artificial seawater was used as the suspending medium to dilute the citrate buffer solution (0.55 M to 0.05 M, unless otherwise mentioned). Once positioned, shear flow was applied at various controlled shear rates to observe the critical conditions for significant unraveling.

\begin{figure*}[!htbp]
    \centering
    \vspace{-1cm}
    \includegraphics[width=\linewidth]{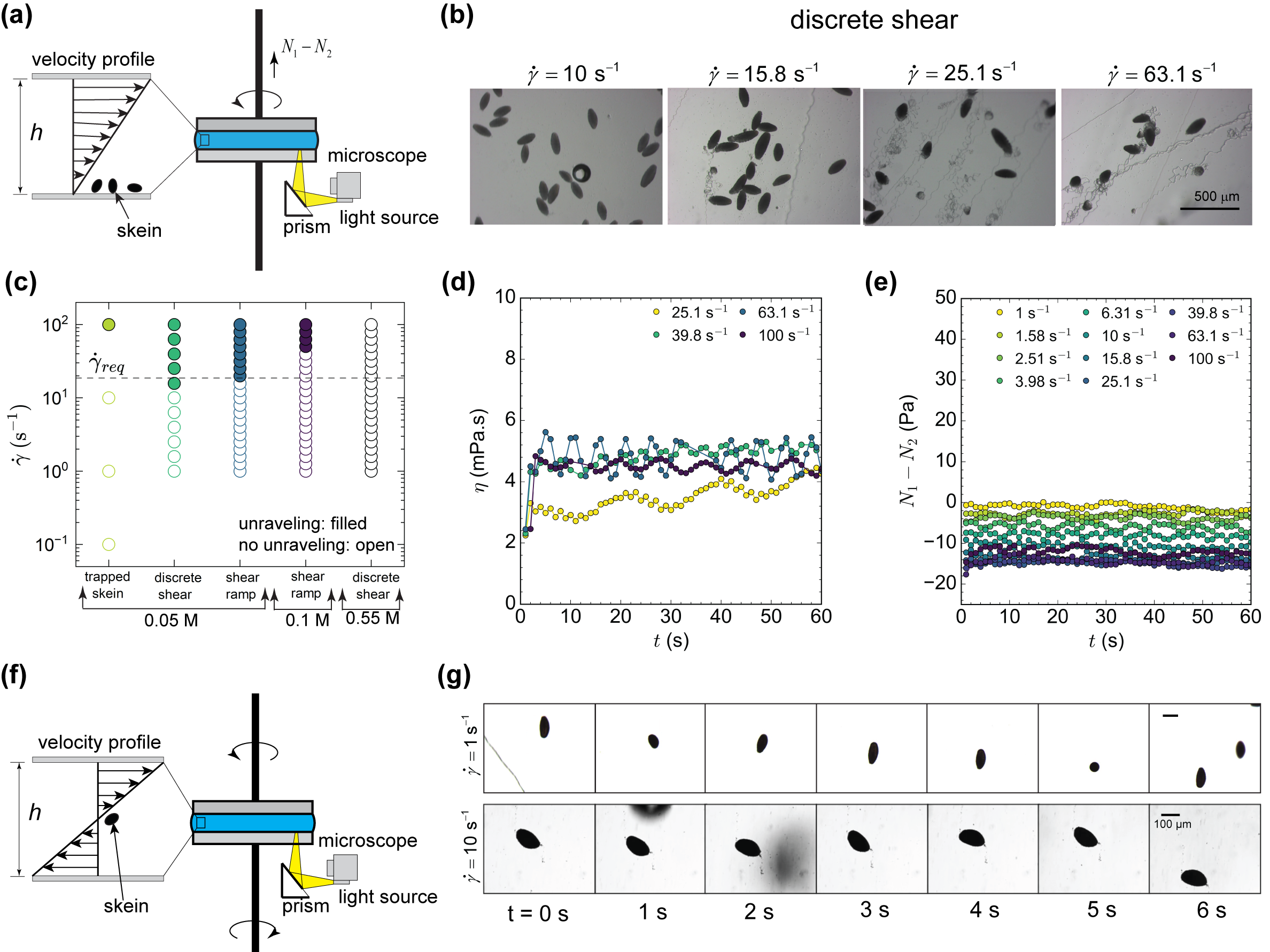}
    \caption{ Characterization of skein unraveling dynamics and rheological behavior under controlled shear flow.
(a) Schematic of rheo-optical setup for multi-skein unraveling under controlled shear conditions. 
(b) Representative images showing partial to full unraveling across different shear rates.
(c) Summary map of unraveling outcomes: filled circles indicate unraveling, open circles indicate no unraveling. Increasing the buffer concentrations suppressed unraveling.
(d) Viscosity evolution during skein unraveling under discrete shear, showing no significant bulk rheological changes after unraveling.
(e) Normal stress difference $N_1-N_2$ over time across all shear rates showing no significant bulk rheological changes after unraveling. (f) Schematic of rheo-optical setup for
positioning isolated skeins in a counter-rotating parallel plate geometry to create a stagnation plane for trapping skeins. (g) time-lapse images of skein behavior under shear flow at two different shear rates: $\dot{\gamma} = 1$~s$^{-1}$ (top row) and $\dot{\gamma} = 10$~s$^{-1}$ (bottom row). No unraveling is observed below $\dot \gamma =10 ~\text{s}^{-1}$.  }
    \label{fig:6}
\end{figure*}

\subsubsection{Unraveling of skeins with mucin vesicles}

We investigated the bulk unraveling behavior of multiple skeins in the presence of mucin vesicles under controlled shear flow, as shown in Figure~\ref{fig:5}a (experimental setup in Figure~S2). Skeins and vesicles were loaded into the rheometer geometry, and the top plate was rotated to impose either a gradually increasing shear rate (ramp) or discrete, constant shear rates. During shear, mucin vesicles observed to deploy in situ, forming a surrounding mucus mass that mimics natural slime deployment. In the ramp protocol, shear rate was increased stepwise (from $\dot\gamma=[1-300]$~s$^{-1}$ with ten points per decade), holding each rate constant for a defined time interval ($t_{\text{ramp}}=[1-5]$~s) before advancing to the next value. For the discrete shear tests, a new batch of skeins with mucin vesicles was used for each shear rate to eliminate flow history effects. Shear rates from $\dot\gamma=[1-100]$~s$^{-1}$ were applied logarithmically, with five discrete values per decade.


Figure~\ref{fig:5}b summarizes the unraveling regime map across a broad range of applied shear rates. Each circle represents an independent trial, with filled symbols indicating skein unraveling and open symbols indicating no unraveling. 
On the left, discrete shear rate experiments reveal a transition from no unraveling to unraveling. The onset of this transition, observed between 10 and 30~s$^{-1}$ (corresponding to a shear stress of 0.03 to 0.1~Pa), is identified based on the first appearance of visibly uncoiling threads extending into the flow. Although not a sharp threshold, this range marks the onset of observable deployment based on optical evidence.
The right portion shows results from ramped shear protocols, which were further subdivided to examine the effects of varying plate gap height ($h=[1-2]~$mm) and ramp time ($t_{\text{ramp}}=[1-5]$~s). 
Despite differences in protocol, the critical shear rate required for unraveling remains largely consistent (gray bounded region). However, we observe a modest decrease in the unraveling $\dot{\gamma}$ at smaller gap heights and a slight increase in the unraveling $\dot{\gamma}$ with faster ramp times. These trends suggest that unraveling is primarily governed by the instantaneous flow strength, though mechanical confinement and deformation rate history can modulate the threshold.
The dashed line denotes the predicted critical shear rate ($\dot{\gamma}_{\text{req}}$) from the unraveling model~\cite{16}, using $F_p=6.8$~nN from the skein peeling force measurements (more details in the discussion).


Figure~\ref{fig:5}c demonstrates the unraveling behavior and rheological response of hagfish skeins with mucus under a continuously increasing shear rate. In this experiment, a constant ramp protocol ($t_{ramp}=5~\text{s}$ and $h=2~$mm) was applied from $\dot{\gamma} = 1$~s$^{-1}$ to $\dot{\gamma} = 100$~s$^{-1}$ over a total duration of 130 seconds. Optical images at each time point capture the progressive deployment of threads from skeins as shear is applied. Minimal unraveling is seen at low shear rates, but as shear increases beyond $\sim$25~s$^{-1}$ ($\sigma=0.08$~Pa), skeins begin to unravel visibly and ultimately form an entangled network by the end of the ramp (Figure~\ref{fig:5}c and Movie S5).

To evaluate whether the unraveling of hagfish skeins affects the macroscopic shear flow rheological properties of the medium, and whether the deployed thread network and mucus mass contribute any measurable change, we measured the shear viscosity ($\eta$) and the normal stress differences in shear ($N_1 - N_2$) during the ramp~\cite{31,32}. 

We defined the reliable operational window by measuring the steady-state shear viscosity of artificial seawater \cite{33}. Figure~S3a shows the reliable steady shear data for artificial seawater $\eta\approx 2$~mPa$\cdot$s was obtained for an intermediate shear rate range of 15-100~s$^{-1}$, which we identify as the valid measurement window. At lower shear rates, the apparent increase in viscosity arises from limitations in the torque resolution ($\sim1~\mu$N$\cdot$m in our case), while at higher shear rates $\dot\gamma>300$~s$^{-1}$, secondary flow affects the measured viscosity, e.g.\ set by critical a Reynolds number, Re$_{\text{crit}}=516$ for a 10\% error in measured to ideal torque~\cite{33}. The apparent normal stress difference can show a downward trend at high shear rate due to inertial artifacts (Figure~S3b) inherent to centripetal acceleration in the rotational parallel-plate geometry, which are corrected (Figure~S3a) for inertia using $(N_1-N_2)_{corr}=(N_1-N_2)+0.2\rho \Omega^2R^2$ where $\rho\approx1000$~kg/m$^3$ is density and $\Omega$ is the angular velocity~\cite{31}. The minimum normal force resolution of the transducer is 0.01~N, which corresponds to an approximate normal stress resolution of $\pm$14~Pa in our setup.

The rheological stress responses (Figure~\ref{fig:5}c) reveal a pronounced increase in normal stress difference ($N_1 - N_2$) at high shear rates. Notably, while the viscosity increases slightly by a factor of 2, a significant positive normal stress emerges, due to the buildup of elastic tension along streamlines (Figure~S4). As we will see later, these rheological responses of the deployed exudate in steady flow are predominantly governed by the mucus matrix itself, rather than the threads.

Figure~\ref{fig:5}d presents results from a discrete shear protocol, where new samples were subjected to a fixed shear rate. Representative optical images show increasing degrees of thread deployment with increasing shear rate, confirming shear-induced unraveling from 25.1~s$^{-1}$ to 63.1~s$^{-1}$ (Movie~S6). The accompanying rheological measurements provide further insights into the dynamic evolution of material properties. The left panel shows a time-resolved plot of the normal stress difference, $N_1 - N_2$, which increases systematically with shear rate and reaches values exceeding 100~Pa at $\dot{\gamma} = 100$~s$^{-1}$, a stark contrast to the negligible normal stress observed in seawater alone. The right panel tracks the apparent viscosity over time, which increases by around 2 times during unraveling. 

To test the hypothesis that prolonged shearing could promote skein unraveling, e.g.\ through enhanced mixing of seawater and buffer, we applied steady shear for 180~s for $\dot\gamma<25.1~\text{s}^{-1}$. If mixing were sufficient to induce unraveling, we would expect gradual thread deployment over time. However, no skein unraveling was observed optically (Movie S7), and both the normal stress difference ($N_1-N_2$) and viscosity ($\eta$) remained constant during prolonged shear for 180~s (Figure~S5). These results contradict the mixing hypothesis, instead suggesting that unraveling requires exceeding a critical shear rate rather than extended mixing alone.

\subsubsection{Unraveling of isolated skeins}

To determine whether mucin vesicles are necessary for flow-triggered unraveling, we conducted additional tests in which only skeins were loaded into the parallel-plate geometry before shear, and the bulk unraveling behavior of multiple skeins subjected to controlled shear flow was investigated, as shown in Figure~\ref{fig:6}a. In this case, a discrete shear rate protocol was applied where a new batch of skeins was loaded into the system to prevent prior flow history from influencing the outcome. Shear rates spanning $1$ to $100$~s$^{-1}$ were applied logarithmically with five discrete values per decade.

Representative images at selected shear rates following 60~s of applied shear are shown in Figure~\ref{fig:6}b. At a shear rate of $\dot{\gamma} = 10$~s$^{-1}$, no unraveling was observed, indicating that the applied fluid stress remained below the threshold required for thread deployment. At $\dot{\gamma} = 15.8$~s$^{-1}$, partial unraveling began to emerge, suggesting proximity to the critical flow strength. As the shear rate increased further to $25.1$ and $63.1$~s$^{-1}$, progressively more skeins transitioned into extended filaments (Movie S8), consistent with the onset of unraveling being governed by a flow-dependent threshold.

Figure~\ref{fig:6}c summarizes these results with an unraveling map. Each point represents a trial, with filled markers indicating observable unraveling and open markers indicating no unraveling. Trials were conducted in three different sodium citrate concentrations (0.05 M, 0.1 M, and 0.55 M), allowing assessment of buffer composition on unraveling response. The data reveal a distinct threshold behavior: unraveling consistently occurs above a critical shear rate of approximately $15$–$20$~s$^{-1}$ (shear stress 0.05--0.06~Pa) for the lowest citrate concentration (0.05 M). However, increasing citrate concentration suppresses unraveling, shifting the threshold to higher shear rates or eliminating it. These results support the hypothesis that unraveling is both shear-rate and environment-dependent, and sensitive to skein stabilization from multivalent anions such as citrate \cite{17,34}.

Figures~\ref{fig:6}d and \ref{fig:6}e show the time evolution of viscosity and normal stress, respectively, for a range of shear rates under discrete shear application. Even at shear rates where visual observations confirmed significant unraveling and extended filaments formation (e.g., $\dot{\gamma} \geq 25$~s$^{-1}$), no measurable change in viscosity was detected. This indicates that unraveled threads did not appreciably alter the bulk flow resistance of the medium. Similarly, the normal stress difference $N_1 - N_2$ remained within the expected baseline attributed to inertial artifacts across all tested conditions. Together, these findings suggest that while unraveling is visually striking and results in the formation of extended filaments, the mechanical contribution of these unraveled threads to the overall shear flow rheology is negligible within the resolution of our experimental setup. 
Our results support the view that the flow rheological properties of hagfish slime are dominated by the mucus matrix, consistent with previous studies showing that both slime and mucus-only preparations effectively clog at extremely dilute concentrations, while unraveled threads alone do not~\cite{19}. This suggests that threads play a minimal role in steady shear flow rheology, but may be crucial in the solid network under deformation modes such as extension or flushing. This aligns with the hypothesis that threads serve to spatially organize and mechanically stabilize the mucus, enabling the formation of a soft elastic network with large internal porosity rather than simply thickening or obstructing flow~\cite{19}.

To investigate whether skein unraveling can occur without surface pinning, we conducted experiments on individual skeins suspended in the stagnation plane between counter-rotating plates. Chaudhary et al.~\cite{16} showed that unraveling driven by viscous drag can occur within a few hundred milliseconds and is accelerated when the skein is pinned against a surface. We aimed to examine whether unraveling can occur for free skeins without pinning. Using a parallel-plate rotational rheometer with counter-rotating plates, we generated a stagnation plane within the flow field, as illustrated in the schematic (Figure~\ref{fig:6}f). While skeins are not naturally buoyant and typically settle near the bottom plate, occasional skeins remained suspended and could be trapped in the stagnation plane by adjusting the relative rotational velocities of the two plates (in most trials, the stagnation plane was located close to the bottom plate). This allowed us to apply defined shear rates while keeping the skein freely suspended, enabling us to isolate the role of flow-induced forces in triggering unraveling without the influence of physical anchoring.

We did not observe skein unraveling under controlled shear flow with an isolated skein trapped in the stagnation plane for shear rate up to $\dot\gamma=10$~s$^{-1}$. 
Figure~\ref{fig:6}g shows time-lapse images of a skein suspended in the stagnation plane of a counter-rotating parallel plate setup under shear rates of $\dot{\gamma} = 1$~s$^{-1}$ (top row) and $\dot{\gamma} = 10$~s$^{-1}$ (bottom row). In both cases, no unraveling was observed over the 6~s duration. Instead, the skein exhibited continuous rotational motion, consistent with classical Jeffery orbits of ellipsoidal particles in shear flow~\cite{35,36}. This tumbling behavior, characteristic of freely suspended anisotropic bodies in viscous flows, highlights that while the skein experiences flow-induced torque, it does not encounter sufficient drag force to initiate unraveling (Movie~S9). Thus, the unraveling does not become any easier in the absence of pinning. However, at shear rates above $10$~s$^{-1}$, skeins were difficult to trap in the stagnation plane due to inertial or hydrodynamic instabilities, limiting precise identification of the critical shear rate in this setup. While unraveling is definitively observed at $100$~s$^{-1}$, the exact threshold for skein deployment lies between $10$ and $100$~s$^{-1}$, consistent with multiple skeins behavior (Figure~S6).

\section{Discussion}

\subsection{Revisiting the fluid mediated unraveling model}

Chaudhary \textit{et al.}~\cite{16} developed a theoretical framework to describe the rapid unraveling of hagfish slime threads under fluid flow. In this model, the unraveling process was governed by the competition between the external fluid traction force and the internal peeling resistance of the skein, characterized by a dimensionless peeling number, $\wp$. When $\wp \gg 1$, the fluid drag greatly exceeds the peeling resistance, and the thread advects with the local flow velocity, resulting in rapid unraveling. The unraveling model was studied in uniform flow and extensional flow, resulting in different flow strengths for unraveling (Figure~S7~\cite{16}). Conversely, if $\wp \ll 1$, unraveling is impeded. The model assumed negligible effects from inertia, filament self-interactions, and external forces such as Brownian motion or gravity, reducing the problem to a balance of hydrodynamic drag and peeling force at the unraveling front. In their formulation, in the absence of direct measurements, the peeling force was modeled as a weakly rate-dependent power-law, $F_p(\dot{L}) = \alpha \dot{L}^m$, where $L$ is the length of the free thread and $0 \leq m \leq 1$. Assuming van der Waals adhesion energies ($E_\text{adh} \sim 50-60$~mJ/m$^2$), they estimated a peeling force on the order of $100$~nN.

Our experimental results revise this critical parameter. Direct \textit{in situ} measurements revealed a significantly lower average peeling force of $F_{p,\text{exp}} = 6.8$~nN, more than an order of magnitude smaller than previous estimates. Reduced peeling resistance suggests that unraveling is even easier than previously hypothesized. 

We observe the skein unravels either when the skein is pinned at the bottom plate of the rheometer, or in flow with free skein and thread.  While the unraveling model is based on extensional flow (pure extension stretches material uniformly, whereas shear combines rotation and stretching), we approximate shear flow using a scaling based on the local velocity $U = \dot{\gamma} R_0$ with $R_0$ being the critical length scale, allowing the peeling number for free skein and thread to be expressed as
\begin{equation} 
\wp \approx \frac{6 \pi \mu \dot{\gamma} R_0^2}{\mathcal{C}}, 
\end{equation}
where $\mu$ is the fluid viscosity, $\dot{\gamma}$ is the shear rate, $R_0$ is the skein radius, and $\mathcal{C}$ is the force resisting peeling. Peeling number $\wp$ as a function of viscosity for different strain rates is plotted in Figure~S8. For seawater viscosity $\mu=2$~mPa.s and skein radius of $R_0=100~\mu$m, we find that $\dot\gamma=18$~s$^{-1}$ ($\dot\gamma=27$~s$^{-1}$ for pinned skein) is required for $\wp=1$, consistent with our observations.

\subsection{Mucus mechanical properties}

The high extensibility of hagfish mucus ($\varepsilon_b=2700\%$) suggests that it may contribute meaningfully to both the formation and mechanical properties of the slime network. Rather than acting as a collection of small elastic junctions, the mucus may undergo large deformations and contribute as a long, stretchable component within the network, enhancing the overall mechanical resilience and deployability of the slime.  While previous studies reported extensional thickening in mucus~\cite{18}, our measurements indicate apparent softening at large strains (Figure~\ref{fig:2}d), though this may reflect geometric assumptions and not the true material response. Future work investigating rate-dependent effects could further elucidate the viscoelastic contributions to the mucus mechanical response.

Figure~S4 shows that hagfish mucus exhibits approximately constant shear viscosity ($\approx 80$~mPa.s) across shear rates at very dilute concentrations, however, still high enough concentrations to show normal stress differences ($N_1-N_2$), similar to Boger fluids (an elastic liquid with a constant viscosity)~\cite{37}. The data was collected by averaging over 10 s as the mucus network breaks down when sheared for longer periods \cite{18}. We obtained significant $N_1-N_2$, which scale approximately as $\sim\dot{\gamma}^2$, consistent with classical predictions for viscoelastic fluids modeled by a constant-viscosity upper-convected Maxwell (UCM) or Oldroyd-B fluid \cite{20}. The presence of large elastic stresses despite Newtonian-like shear viscosity highlights the complex rheology of mucus.

Figure~S9 shows the evolution of the normal stress difference ($N_1-N_2$) and viscosity ($\eta$) during mucin vesicle deployment under different applied shear rates. The buildup of $N_1-N_2$ and $\eta$ follows similar trends both with and without skeins, suggesting that the primary source of the complex, time-dependent bulk flow stress response during unraveling originates from the mucus matrix rather than the threads. 

\subsection{Localized coiling structure and its impact on peeling forces}

Our peeling force measurements (Figure~\ref{fig:3}) reveal a distinctive sawtooth pattern, characterized by sequential force drops and recoveries during thread extraction (silk fiber composites produced by spiders show a similar force extension profile due to the loops unraveling~\cite{38}. This signature is consistent with a hidden length mechanism, where initially inaccessible portions of the thread must be thwarted and unthwarted before being pulled free, rather than a smooth, continuous peeling process. Such behavior suggests that thread deployment involves overcoming discrete mechanical barriers, potentially linked to the non-uniform coiling architecture within the skein. Recent 3D reconstructions of thread skeins \cite{8} suggested the conical, nested loop structures and circumferential cabling along the skein surface imply complex geometric constraints that vary across the skein, especially between the skein tip and its deeper interior. Therefore, the peeling force is not a simple function of drag versus adhesion but also reflects the local topology of the coiled threads, offering a mechanical basis for the punctuated unraveling dynamics observed during hagfish slime deployment.




\subsection{Other considerations}

Our measured peeling forces suggest that thread deformation during peeling is confined to the elastic part of the slime thread stress-strain curve, offering insight into the eventual collapse of the slime network over time. For a peeling force of \( F = 6.8 \, \text{nN} \) acting on a thread with diameter \( d = 2 \, \mu\text{m} \), under the assumption of uniaxial tension we have $\epsilon  \approx$ 0.00034. 
This corresponds to a strain of approximately 0.034\%, which is substantially below the reported yield strain of hagfish threads (\( \epsilon_{\text{yield}} \approx 0.34 \))~\cite{15}. However, in the chaotic flow conditions of a predatory attack, local forces could exceed this threshold, potentially causing plastic deformation behavior. Yet, the threads may retain some elastic memory with a tendency to recoil after deployment.  Over time, such recoil, combined with mucus drainage or mucin aggregation, may lead to partial collapse of the slime network \cite{13,18,19}.

This study has several limitations that should be considered when interpreting the results. First, all peeling force measurements were conducted in sodium citrate buffer, which, while effective for stabilizing the exudate, is non-physiological and may not fully capture the ionic conditions or mucin-swelling behavior present in natural seawater. As such, the adhesive interactions between threads and mucus could differ from \textit{in vivo} conditions, and the use of buffer likely leads to an overestimation of the peeling forces required under natural conditions. We observed skeins unraveled at higher shear rate for increased buffer concentrations (Figure~\ref{fig:4}c), and the magnitude of shear normal stress differences is higher at lower buffer concentrations for the same shear rate of $\dot\gamma=63.1~\text{s}^{-1}$ (Figure~S10). Additionally, we did not measure mucin concentration nor directly characterize the viscoelastic response of the mucus. Rheological measurements were performed using a parallel-plate geometry, where the local shear rate is not uniform but varies linearly with radial position as $\dot{\gamma}(r) = \Omega r/h$ where $\Omega$ is the rotational speed, $r$ is the radial distance from the center and $h$ is the gap between parallel plates. Consequently, only peripheral regions experience the nominal reported shear rate, and unraveling likely initiates near the high-shear periphery.

\section{Conclusion}

In this study, we present the first quantitative measurements of the micromechanical forces governing the unraveling and deployment of hagfish slime threads. Using a custom micromanipulation setup and rheo-optical characterization, we measured the peeling force required to liberate threads from skeins, adhesive interactions with mucus, and cohesive forces between threads. Our results show that the median peeling force is approximately 6.8~nN in these conditions, significantly lower than previous estimates, and that thread–mucus adhesion is substantially stronger than thread–thread cohesion.

Incorporating these \textit{in situ} force measurements into a peeling-based unraveling model predicts a critical flow strength (18 s$^{-1}$) required for skein deployment that is consistent with our experimental observations of \( \dot{\gamma} \approx 15-25~\mathrm{s}^{-1} \) ($\sigma=$0.05 Pa to 0.08 Pa). 
This shear rate range is comparable to the near-wall conditions in moderate laminar flow of water through narrow tubes (diameter $\sim$3~mm) at average velocities of 6–9~mm/s, offering an intuitive reference for the magnitude of flow required for slime deployment.
Despite the visual transformation associated with thread deployment, bulk rheological flow stress, i.e.\ viscosity and normal stress differences, remained largely unchanged when threads unfurled, highlighting the non-intuitive contribution of the thread network in dilute slime.

Our findings support a mechanistic view where unraveling is driven by fluid-induced forces that overcome localized peeling resistance, with mucus acting as a viscoelastic matrix that supports but does not fundamentally alter the deployment threshold. The thread elasticity, mucus extensibility, and negligible adhesion between threads together enable rapid, scalable formation of a soft yet structurally coherent defensive network. These insights not only resolve key uncertainties in the biophysics of hagfish slime but also inform future designs of synthetic deployable fiber–gel systems.


\section{Acknowledgements}
This work is supported by Defense Advanced Research Project Agency (DARPA), under contract no. N660012124036. R.H.E. acknowledges Anton Paar for providing the MCR~702 rheometer used in some of the rheological experiments.


\end{document}